# Momentum exchange in the electron double-slit experiment.

H. BATELAAN, ERIC JONES, WAYNE CHENG-WEI HUANG, ROGER BACH

*Department of Physics and Astronomy, University of Nebraska—Lincoln, 208 Jorgensen Hall, Lincoln, Nebraska 68588-0299, USA*

PACS 03.65.Ta – Foundations of quantum mechanics; measurement theory

PACS 03.75.-b – Matter waves

PACS 03.65.Vf – Phases: geometric; dynamic or topological

**Abstract.** We provide support for the claim that momentum is conserved for individual events in the electron double slit experiment. The natural consequence is that a physical mechanism is responsible for this momentum exchange, but that even if the fundamental mechanism is known for electron crystal diffraction and the Kapitza-Dirac effect, it is unknown for electron diffraction from nano-fabricated double slits. Work towards a proposed explanation in terms of particle trajectories affected by a vacuum field is discussed. The contentious use of trajectories is discussed within the context of oil droplet analogues of double slit diffraction.

**1. Introduction.** Recently we performed the electron double slit experiment, and the pattern was recorded one electron at-a-time [1]. The electron detection rate was about one electron per second. This made it possible to manually turn off the electron source after the first electron was recorded. This electron can, by chance, land in a first diffraction order (see Fig.1). This can be considered a completed single-event experiment. Often single events experiments are only considered in a probabilistic way as the best theory available to compare with, that is Quantum Mechanics, is probabilistic. Nevertheless, a quantum description also includes the correct prediction that the individual, in this case position, outcomes are eigenvalues of operators. Even more is known about single events. This becomes clear upon asking the question: "Is momentum conserved for this experiment?" We will provide support for the claim that the generally accepted answer is yes. The natural follow-up question that is the central theme of this paper is: "By what mechanism do the electron and the slit exchange momentum?" We claim that the answer is not known and that the question is a valid one. Some discussion on possible mechanism is given. In particular, the role of image charge interaction between the electron in double slit walls and the vacuum field is discussed. The proposed explanation that the double slit provides a boundary condition for the vacuum field, which in turn provides a means by which the electron trajectory exchanges momentum[2–6] with the slit is discussed within the context of the theory Stochastic Electrodynamics (SED) [2,7]. The provocative possibility of any trajectory explanation is considered in view of the well-known oil-droplet double slit analogue. The validity range of SED and the relation with the Heisenberg uncertainty relation are discussed for the Harmonic oscillator. The intent of this paper is to raise questions and discuss ongoing work that is unfinished and as of yet inconclusive.

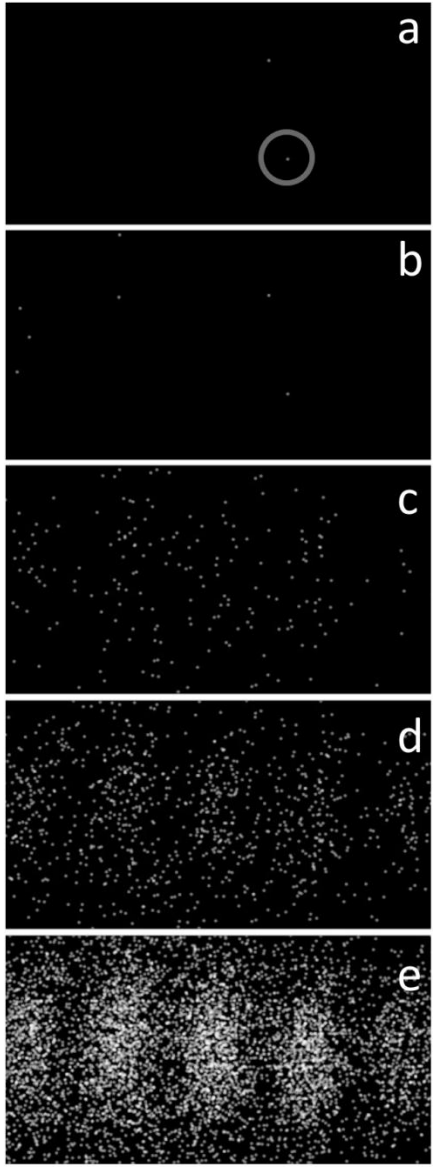

**Figure 1: Buildup of Electron Diffraction.** "Blobs" indicate the locations of detected electrons. Shown are intermediate build-up patterns from the central five orders of the diffraction pattern, with 2, 7, 209, 1004, and 6235 electrons detected (a-e) [1].**The circle indicates where the first electron landed for this data run.**

**2. Momentum conservation in double slit diffraction.** The famous Einstein-Bohr dialogue sheds light on the question of the momentum conservation [8]. Einstein attempted to prove that quantum mechanics is incomplete. In this famous series of discussions between Einstein and Bohr, several topics in quantum mechanics were debated.

One of these is the double slit interference experiment [9,10]. Quantum mechanics predicts an interference pattern with perfect contrast, but only if we have no knowledge of which slit the particle went through. Einstein devised a thought-experiment that would measure through which slit the particle went *and* show the interference pattern. If such an experiment could be performed he hoped that it would show that quantum mechanics is incomplete. Einstein considered a particle beam illuminating a single slit screen placed in front of a double slit screen. Ingeniously he suggested that the recoil of the single slit screen could be measured to determine through which slit the particle would move. Einstein used momentum conservation to predict the recoil. Because the recoil would be present regardless if it is measured or not, it appears that such a measurement would not affect the experiment in any way, and the interference pattern would remain. Bohr replied that when we measure the recoil, i.e. the momentum of the screen, accurately enough to determine through which slit the particle went, the uncertainty in our knowledge of the position of the slit is so large that the interference pattern is obscured. In other words, quantum mechanics, through Heisenberg's uncertainty relation, protects itself. Also Bohr assumed that momentum conservation holds. In a later treatise Wootters and Zurek analyze this thought experiment quantitatively [11]. This allows setting up a quantitative relation between the probability of going through one slit with the contrast of the interference pattern.

First we will summarize the main results of Wootters and Zurek's approach, using the same notation as used by these authors. The system discussed is schematically presented in figure 2. The position of the single slit plate is given by $z$, the position of the particle at the detection screen is given by $\xi$, the positions of each of the double slits of labeled with A and B and the observed position distribution is indicated by $f(\xi)$.

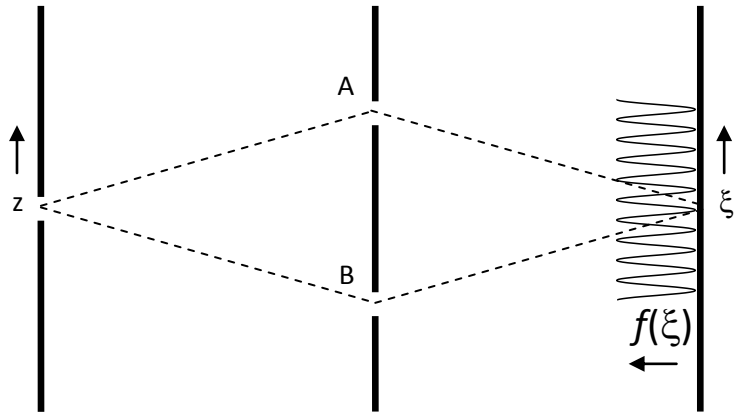

**Figure 2. Schematic of Zurek's analysis of the Bohr-Einstein double-slit thought experiment.**

The wavefunction describing both the slit and particle in position representation is given by

$$\psi(z,\xi) = \frac{1}{\sqrt{2a}\pi^{1/4}} \int e^{-x^2/2a^2} [e^{ik_0(\xi+x)} + e^{-ik_0(\xi+x)}]\delta(z-x)dx. \qquad 1$$

The momentum of the particle is given by $\hbar k_0$, and the parameter $a$ is a measure of the width of the Gaussian distribution of the slit. If $a=0$ then we know where the slit is, but we do not know what its momentum is, and the interference pattern has full contrast. Einstein's modification of the double slit experiment means we would know something about the momentum of the slit, which, as Bohr pointed out, would smear out the position of the slit, because of Heisenberg's uncertainty principle, and thus cause the interference contrast to disappear. Wootter and Zurek's equation 1.1 allows that statement to be made quantitative. For a given value of $a$, the interference pattern can be calculated;

$$f(\xi) = \int |\psi(z,\xi)|^2 dz , \qquad 2$$

by integrating over the unobserved slit position z. Specifically,

$$\begin{aligned} f(\xi) &= \int |\psi(z,\xi)|^2 dz = \\ &= \int \frac{1}{2a\sqrt{\pi}} e^{-z^2/a^2} (e^{ik_0(\xi+z)} + e^{-ik_0(\xi+z)})(e^{-ik_0(\xi+z)} + e^{ik_0(\xi+z)}) = \\ &= \int \frac{1}{a\sqrt{\pi}} e^{-z^2/a^2} (1+\cos(2k_0(\xi+z))) = \\ &= 1 + e^{-a^2 k_0^2} \cos(2k_0 \xi) \end{aligned} \qquad 3$$

In the first step the integration over the delta-function, $\delta(z-x)$, has been performed, while in the last step the integral, $\int \frac{1}{a\sqrt{\pi}} e^{-z^2/a^2} \cos(2k_0 z) = e^{-a^2 k_0^2}$, has been used. In the limit of $a \to 0$ the position of the slit is exactly know, and the contrast is maximum. We now turn our attention to the wavefunction of the slit and the particle in the momentum representation. Wootters and Zurek state that it can be verified that this is "equal" to the position distribution, and give

$$\psi(k_0,\xi) = \frac{a}{2\pi^{1/4}} \int D^{1/2}(\kappa)[p_A^{1/2}(\kappa)e^{ik_0\xi} + p_B^{1/2}(\kappa)e^{-ik_0\xi}]e^{i\kappa z} d\kappa . \qquad 4$$

In this expression, $\kappa$, is slit momentum, while the functions $D$ and $p_{A,B}$ are given by,

$$\begin{aligned} D(\kappa) &= \frac{a}{2\sqrt{\pi}} \left\{ e^{-a^2(\kappa+k_0)^2} + e^{-a^2(\kappa-k_0)^2} \right\} \\ p_A/p_B &= e^{-a^2(\kappa+k_0)^2} / e^{-a^2(\kappa-k_0)^2} ; p_A + p_B = 1 \end{aligned} \qquad 5$$

The functions $p_{A,B}$ give the probability of going through slit A or B, respectively, and $D$ normalizes the wavefunction. We note that the limit $a \to \infty$ justifies the statement in the introduction that diffraction essentially creates and entangled state. In this limit the probabilities act as delta-functions, $\delta(\kappa+k_0), \delta(\kappa-k_0)$. After integration the wavefunction becomes

$$\lim_{a \to \infty} \psi(k_0, \xi) = \tfrac{1}{2}\sqrt{2}(e^{ik_0\xi}e^{-ik_0 z} + e^{-ik_0\xi}e^{ik_0 z}) =$$
$$= |\psi(-k_0)\rangle_{slit} |\psi(k_0)\rangle_{photon} + |\psi(k_0)\rangle_{photon} |\psi(-k_0)\rangle_{slit}$$

We can now proceed and calculate the probability to find the photon on position $\xi$ on the detection screen,

$$f(\xi) = \int \lim_{a \to \infty} |\psi(z,\xi)|^2 dz =$$
$$= \lim_{a \to \infty} \frac{1}{(2a)^2} \int_{-a}^{a} 1 + \cos(2k_0(\xi - z)) dz = 1/2a$$

In words, this states that, if $a$ is made very large then the momentum of the slit is known, and at the same time the interference pattern is known. Combining this with the earlier statement that for very small a the position of the slit is known, while at the same time the interference pattern has full contrast, it can be recognized that this result can be reached by using Bohr's argumentation using the Heisenberg uncertainty relation. Specifically, if the momentum of the slit is measured exactly the position is completely unknown and incoherently averaging over this position blurs the interference pattern completely. Wootters and Zurek not only justified this reasoning, but also give quantitative expressions when not considering the extreme cases. The probability to go through one of the holes and the interference contrast can be calculated for arbitrary values of a.

For this paper, the second line in equation 6 is relevant. It expresses that momentum conservation between the electron and the slit holds. This is used to set up the main question posed in this paper: "By what process does the electron exchange momentum with the double slit?"

**3. Momentum exchange mechanisms**. Upon asking this question in Physics Colloquia (presented by HB), the answer provided by physics professors is surprisingly varied. Answers range from: "the electron induced an image charge in the double slit which back-acts on the electron," and "the electron excites phonons in the double slit, which back-acts on the electron," to "the vacuum field bounded by the double slit structure, acts on the electron." Additional to such answers, the comment is often made that if one does entertain this question, one should not forget that atoms, photons, and neutron all diffract and thus the mechanism should have some rather ubiquitous elements in it. About half of the comments made, support the idea that this is a question that one should not ask or is already answered by the presence of a potential that describes the double slit structure.

To address the comment, whether or not we should ask the question by what process the electron exchanges momentum with the slit, let's consider the same question for the Kapitza-Dirac effect [12].

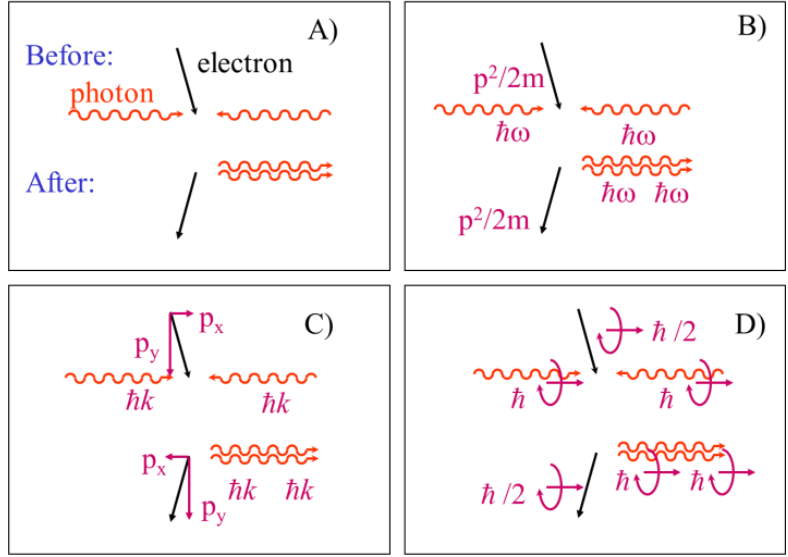

**Figure 3. Kapitza-Dirac effect. Electrons can be diffracted from a standing wave of light** [13]. **A) Stimulated Compton scattering is the basis for the Kapitza-Dirac effect. B) Energy is conserved in this process. C) Momentum is conserved in this process. D) Angular momentum is conserved when the electron spin does not flip.**

When electrons pass through a standing wave of light with periodicity $d \approx \lambda/2$, it is possible for the electron wave to diffract from the periodic light structure with a diffraction angle $\theta \approx \lambda_e/d$, where $\lambda_e = 2\pi\hbar/p_e$, and $p_e$ is the electron's momentum. This effect, known as the Kapitza-Dirac effect, was proposed in 1933 [4] and we realized this experiment in 2001 [13,14]. The process by which the electron exchanges momentum with light is stimulated Compton scattering. One photon is absorbed, while the emission of another is stimulated (Fig. 2A). Energy and angular momentum are conserved in this process (Fig. 2B, 2D). As the absorption and stimulated emission are due to photons coming from opposite directions, the electron experiences a recoil of $2\hbar k$ momentum (Fig. 2C), where $k = 2\pi/\lambda$. At a basic level it is easy to verify that the scattering angle $\theta \approx 2\hbar k/p_e$ and the diffraction angle are identical supporting the explanation of electron diffraction by a "light grating" as stimulated Compton scattering. At a more formal level, perturbation theory and second quantization of the light field can be used to support this claim [15]. The understanding of the mechanism also leads to predictions. When the polarization of the counter propagating light beams is chosen to be perpendicular, no standing wave forms and the electrons do not diffract. Or in the particle picture; angular momentum conservation does not work for photons that carry opposite angular momentum. Before the interaction the two photons carry a total of zero angular momentum, while after the stimulated emission the two photons carry two units of $\hbar$ angular momentum. The electron can at most change its angular momentum by one unit of $\hbar$ in a spin flip process [15].

Inspection of the electron diffraction pattern from light, and from the double-slit reveal, not surprisingly, a very similar phenomenology (Fig. 3). And, a standard quantum mechanical description of the experiments gives good agreement in both these cases. The surprise is that the mechanism can be explained for a light-grating, but not the double-slit case. Let's consider electron diffraction from an ionic crystalline lattice as in the famous Davisson-Germer experiment [16]. Can in this case the mechanism for momentum exchange be explained? An electron experiences the periodic Coulomb potential of the ionic lattice. Coulomb scattering at the particle level is understood as the scattering of virtual photons. Thus the basic mechanism for momentum exchange between the diffracting electron and the ionic lattice is understood. Additionally, predictions can be made. As the lattice heats and the potential shape is modified due to the averaging of the motion of the ionic lattice, the diffraction pattern is modified.

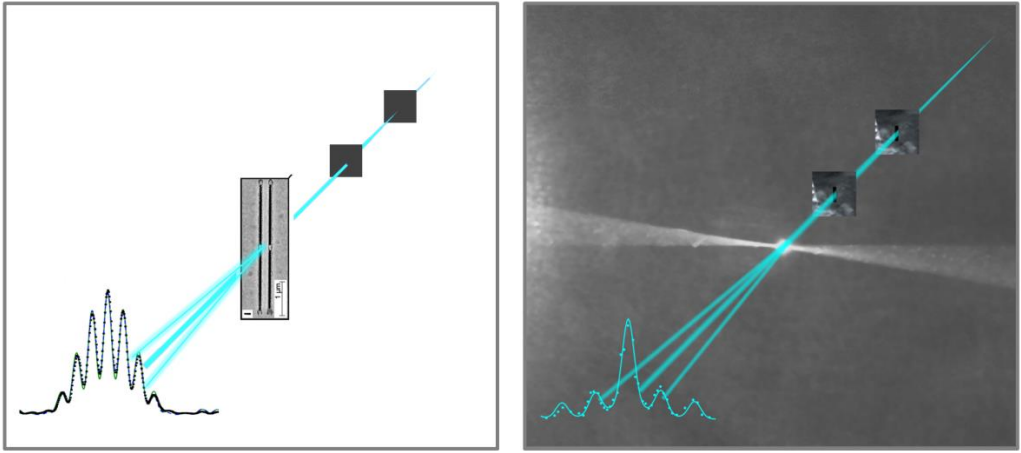

**Figure 4. Electron double-slit and Kaptiza-Dirac diffraction. Left) Electrons (blue lines, starting right top) pass through a nanofabricated double slit (distance between the slits is 300 nm, and shown is an electron microscope image). The measured diffraction pattern (dots) agrees with a quantum mechanical analysis (lines). Right) Electrons (blue lines, starting right top) pass through a standing wave of light (periodicity 266 nm, and shown is a photograph of the laser focus made visible with smoke). The measured diffraction pattern (dots) agrees with a quantum mechanical analysis (lines).**

Is the electron diffraction by a nano-fabricated grating [17] a case that is very similar in nature to the ionic crystal, in that simply the lattice constant is much larger? The ionic crystal lattice has a priodicity of about 2 Angstrom, while the grating has about 1000 Angstrom. Even though the periodicity is very different, the physics appears similar. The electrons are blocked by the grating bars by scattering of the material that the bars are made of. This is due to the Coulomb interaction with the ionic lattice of the material. However, in another sense the ionic crystal is much more like the light grating. For both the light grating and the ionic lattice, the electron experiences a phase grating. That is, the electron wavefunction accumulates a phase that is dependent on space. For the nanofabricated grating the electron experiences an amplitude grating . That is, the electron wavefunction experiences a modulation of its amplitude as a function of position. Additionally the grating bars have no charge or other field in between the grating bars.

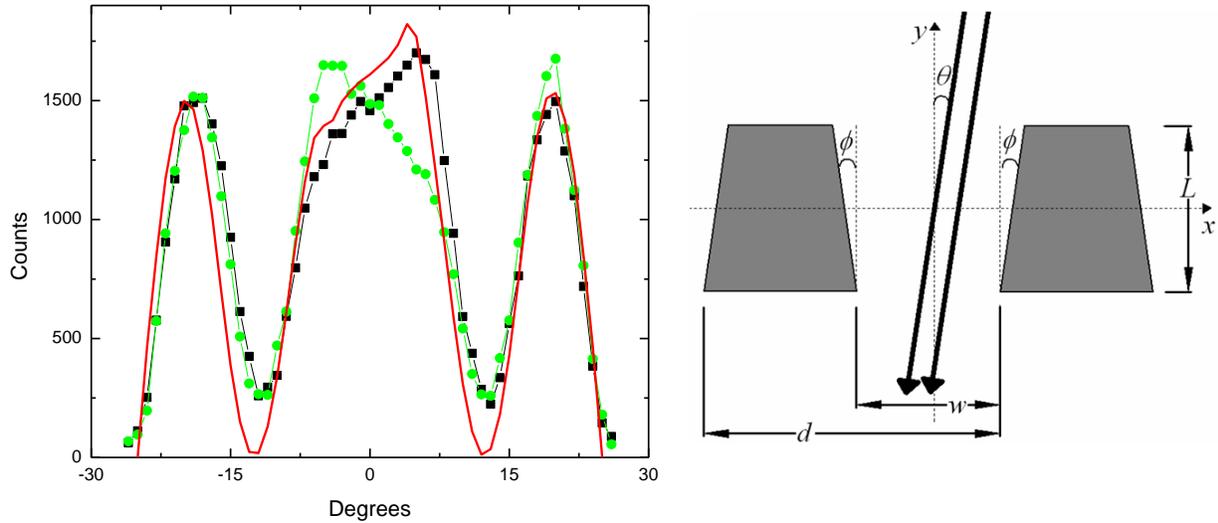

**Figure 5. Image charge effect. An electron beam is diffracted from a nanofabricated grating** [17,18]. **The electron transmission into the third diffraction order is given as a function of grating tilt angle (left). The squares and circles are data for the positive and negative third order. The solid line is the result of a path integral calculation for the positive third order with an image charge potential and the geometry of the grating included. The geometry is given by the cross-sectional cut of the nanofabricated grating (right ). The bevel angle of the slit is given by** $\phi$ **, the electron beam angle with respect to the grating is** $\theta$ **, the slit width is given as *w*, and the grating thickness is given by** $L$ **.**

Returning to the double slit, it is analogous to the nanofabricated grating, but instead of many slits it has only two. One of the suggestions for a mechanism for electron diffraction is that the electron would induce an image charge in the double slit material walls, which in turn would back-act on the electron. This would, after all, provide real fields, and thus lead to the exchange of virtual photons with the electron within the slits. This has been investigated experimentally. The image charge interaction has been measured to weakly modify the diffraction pattern [18]. The slope on the diffraction rocking curve around $\theta = 0$ is due to the image charge. In the absence of image charge this slope is zero. The basic idea is that electrons that move with angle where $\theta \approx \phi$ (see Fig.5 right) are closer to one grating wall and thus experience an image charge potential gradient, modifying the diffraction pattern. The removal of the image charge term in the theoretical treatment leaves the diffraction intact, thus the image charge does not explain electron diffraction.

Another suggestion is that the electron would excite phonons in the double slit material. This does not directly identify the mechanism of diffraction, but would be specific about what two objects interact and exchange momentum. It would be the electron and a phonon. This idea has never been tested, but may lead to such ideas as exciting phonons in the double slit material before the electron diffracts and thereby controlling the electron diffraction pattern.

Another idea is that the double-slit poses a boundary condition on the vacuum field and the modified vacuum field affects the motion of the electron. This may appear akin to a Casimir effect, but this is not what it is. The leading term of the vacuum energy interaction of an electron and a wall is the image charge [19,20]. The QED corrections to that term are very small and do not explain diffraction. In Stochastic Electrodynamics the vacuum field interacts with a charged point particle. This theory has had some successes in for example correctly obtaining Casimir forces [21] and the absorption spectrum of the harmonic oscillator [22]. For an extensive recent review see [23]. On page 323 of this book a bundle of trajectories is shown for the double slit (Fig. 6). No probability distribution has been calculated yet. It is perhaps surprising that a theory that supports real trajectories is entertained to provide insight on a physics phenomenon that is considered to be a hallmark of quantum mechanics. Quantum mechanics is not consistent with non local-real theories as evidenced by experimental tests of Bell's inequalities [24–26]. The notion of trajectories in this context leads inevitably to discussion.

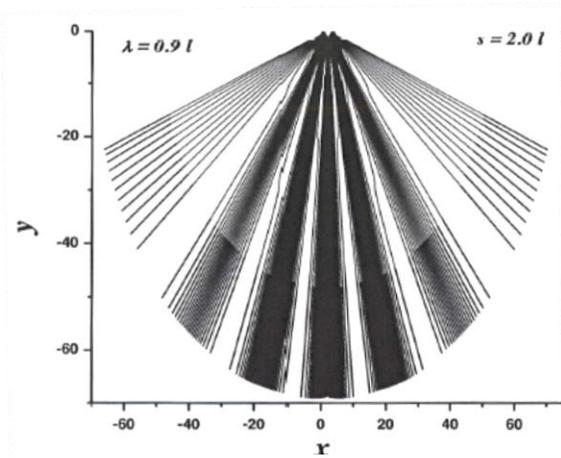

**Figure 6. SED trajectory simulation. Trajectories are shown for a simulation of electrons for the double slit experiment. Figure courtesy of J. Avedano.**

**4. On the possibility of trajectory explanations.**
Recently, there has been a report on oil-droplets that show a behavior that is analogous to electron diffraction from a double slit [27]. The diffraction pattern has been recorded one droplet at-a-time, and the oil droplet trajectories have been recorded. This experiment and its interpretation appear at odds with many claims that have been made on double slit diffraction. For example, Feynman stated: "We choose to examine a phenomenon which is impossible, *absolutely* impossible, to explain in any classical way, and which has in it the heart of quantum mechanics. In reality, it contains the only mystery." [10] As the size of the oil droplet is about 0.8 mm, it is clearly not isolated sufficiently from its environment to require a quantum mechanical description. A classical description is completely sufficient. The oil droplet report should thus be considered to describe a remarkable phenomenon that requires extensive scrutiny. Other research groups are attempting to repeat the oil droplet experiment [28]. In particular, the group of Bohr claims: "that the single-particle statistics in such an experiment will be fundamentally different from the single-particle statistics of quantum mechanics" [29]. Our group is also in the process of repeating the oil droplet experiment. Single- and double slit diffraction was studied. The experimental set-up follows the design of the Couder group and a short description of the experiment is given below for completeness.

A flat, horizontal square dish is filled with approximately 4 mm of oil. The dish oscillates in the vertical direction with a frequency of 50 Hz, which allows a droplet of about 0.8 mm size to bounce on the oil surface for periods up to hours. Just below the Faraday threshold for surface excitation, the oil droplet and the mostly circularly shaped wave that it excites, move together at a constant velocity. This association of a droplet and a wave has been called a "walker." The droplet surfs the wave and the wave guides the droplet motion. As the wave is modified when it moves in the vicinity of a physical boundary, it steers the droplet. The physical boundary used in this experiment is a submerged slit structure. A bundle of trajectories is shown in Fig. 7.

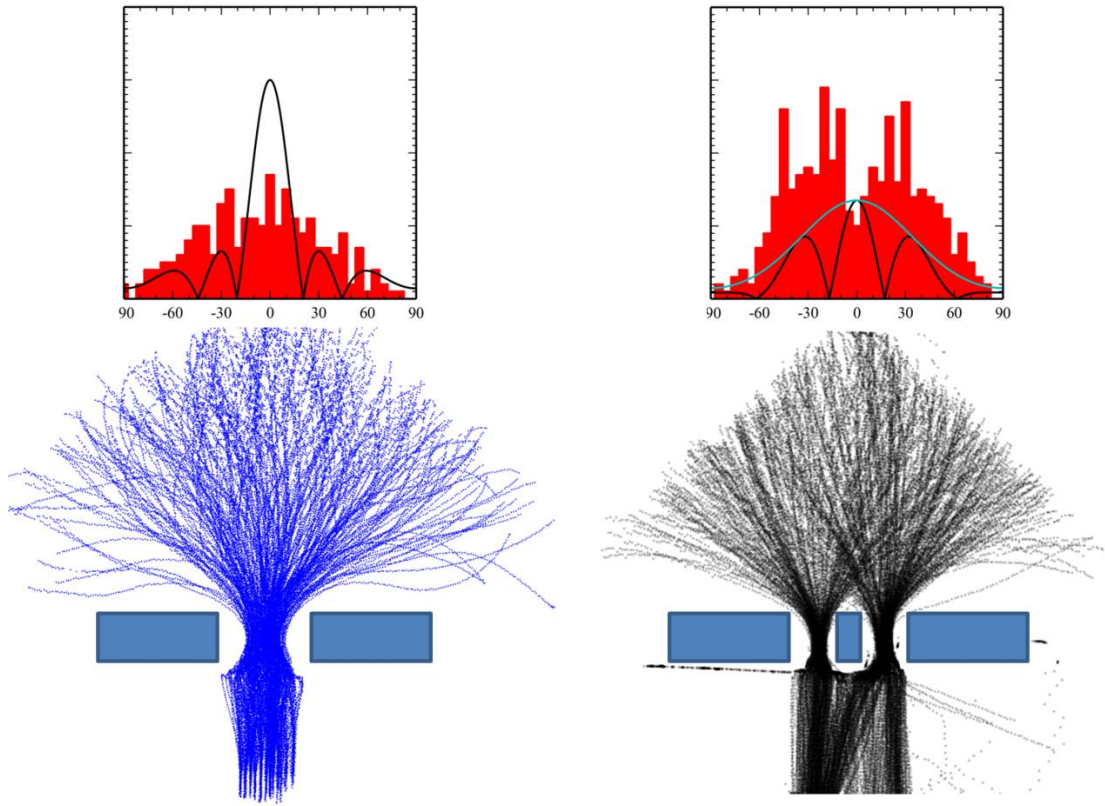

**Figure 7. Oil droplet diffraction analogue. Oil droplet trajectories for a single slit (left bottom) and for a double slit (right bottom) measured at UNL are shown. The corresponding angular trajectory distributions are given in the top graphs. The viscosity used for the single and double slit is 20 cSt and 50 cSt, respectively. The solid line is the fit given by Couder to the data reported in [27]. For a description see text.**

The trajectories are recorded with a web camera. Data analysis software finds the position of the droplet at regular intervals and a trajectory is built as a series of timed positions. In Fig.7 the probability distribution of trajectory angles, after the droplet has passed the slit, has been given. The angles are found at a distance of two slit widths (measured from the center of the slit). The angular distribution does not have distinct peaks. The angles for the single slit case, at which the peaks should occur are approximately given by $\theta \approx \lambda_F / d$, where $\lambda_F$ is the Faraday wavelength of the oil bath waves and $d$ is the width of the slit. These are the locations of the maxima in the solid curve. The angles observed further from the slit are affected by the walls. The idea is that the extent of the oil waves should be sufficient to interact with the entire slit, but smaller than the square walls of the bath, so that the oil droplet's motion is affected by the slit and not the walls. However, after the oil droplet has passed through the slit, the oil waves will interact with the wall and change the droplet motion. For the double slit an example is given where a double lobed distribution is found in contrast with the results reported by Couder. However, this data is for one particular fluid depth and shaking amplitude of the oil bath. Additionally, the initial angular distribution is not post-selected in the double slit example shown. In analogy with an electron diffraction experiment, it is important to limit the divergence of the incoming particle "beam" to less than the diffraction angle. The initial position distribution should be flat to represent a beam with a uniform distribution. The final angular distribution depends strongly on these choices and values. Figure 8 shows another single slit run for slightly different parameters (that is symmetrized following the approach by Couder) that exhibits diffraction like peaks at approximately the correct angle. At this time we cannot draw general conclusions due to the limited amount of data available.

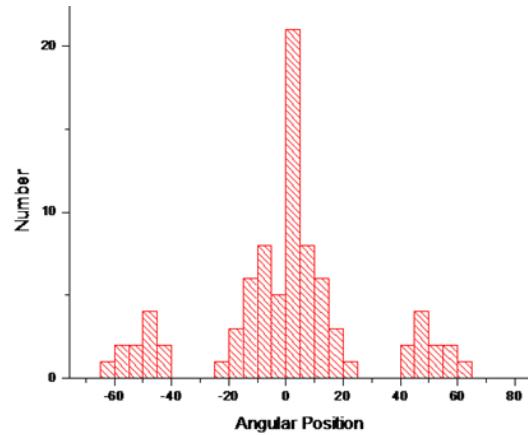

**Figure 8. Single-slit oil droplet distribution. The angular distribution shows diffraction-like peaks (oil viscosity of 50 cSt).**

The interpretation of the thought-provoking experiments by Couder, have also been reexamined by Bohr. Bohr has repeated the double slit experiments and does not find diffraction peaks. Bohr's analysis leads to the claim that the oil droplet experiments do not have the capability to show interference-like behavior. It appears that the data provided by Bohr does not use an initial uniform distribution. It is anticipated that more refined results from multiple research groups are forthcoming and may shed light on the veracity of the oil droplet experiment and the meaning of the analogy. Are then explanations of double slit diffraction using trajectories even possible? If the Couder oil droplet experiments find some confirmation this question could be affirmatively answered.

**5. The role of the vacuum field.** Returning to the proposed vacuum field explanation of electron diffraction, we decided to study the theory of SED for the harmonic oscillator for the purpose of establishing its validity regime. The motivation is that SED yields the ground state of the harmonic oscillator which obeys the Heisenberg uncertainty relation. The Heisenberg relation is central to diffraction. A simple argument can illustrate this. In electron diffraction from a single slit of width $d$, the extent of diffraction angles $\theta$ obeys the diffraction equation $\theta = \lambda_{dB}/d$ for small angles. On the other hand the extent of the diffraction angles are an indication of the momentum uncertainty in the direction of the slit $\theta = \Delta p_x / p$. Using the de Broglie wavelength $\lambda_{dB} = h/p$, equating the slit width with the uncertainty of the position $d = \Delta x$, the uncertainty relation $\Delta x \Delta p_x = h$ is now recovered. As SED gives a model explanation for the ground states of the harmonic oscillator, it appears natural to suspect that it can also explain diffraction.

Boyer has shown that the moments $\langle x^n \rangle$ of a harmonic oscillator immersed in the SED vacuum field are identical to those of the quantum harmonic oscillator in the ground state [8]. As a consequence, the Heisenberg minimum uncertainty relation is satisfied for an SED harmonic oscillator, and the probability distributions of the SED harmonic oscillator is also the same as that of the ground state quantum harmonic oscillator. In [30] we ask the question: "What happens to the dynamics of the classical harmonic oscillator in the presence of the SED vacuum field such that its probability distributions would become Gaussian?" and "Why do the widths of these distributions satisfy Heisenberg's minimum uncertainty relation?" The answer is that the vacuum field drives the electron motion while radiation damps it. The energy balance leads to an average energy for the particle of $\hbar\omega/2$. Planck's constant enters through the overall strength of the vacuum field and its value is determined by experiment. The vacuum field is found as a Lorentz boost invariant solution of the Maxwell equations in free space [31]. In this sense, which is not widely recognized, SED is a theory that is independent of quantum mechanics.

The vacuum modes that are used, cover the damped harmonic oscillator resonance width. No other vacuum modes are used in the simulation. The results converge as the width of the spectrum is increased. The vacuum field as a function of time can be approximated by a single mode for less than a coherence time $\tau$, which is the reciprocal of the vacuum field bandwidth. The particle's response to this field is that of a damped harmonic oscillator driven by one frequency and has the usual double peaked classical probability distribution. Averaging over the particle motion over many coherence times leads to a Gaussian probability distribution that agrees with the Quantum mechanical distribution (Fig. 9). Also, the product of the widths of the momentum and position distributions satisfies the equality in Heisenberg uncertainty relation in accordance with a minimum uncertainty packet.

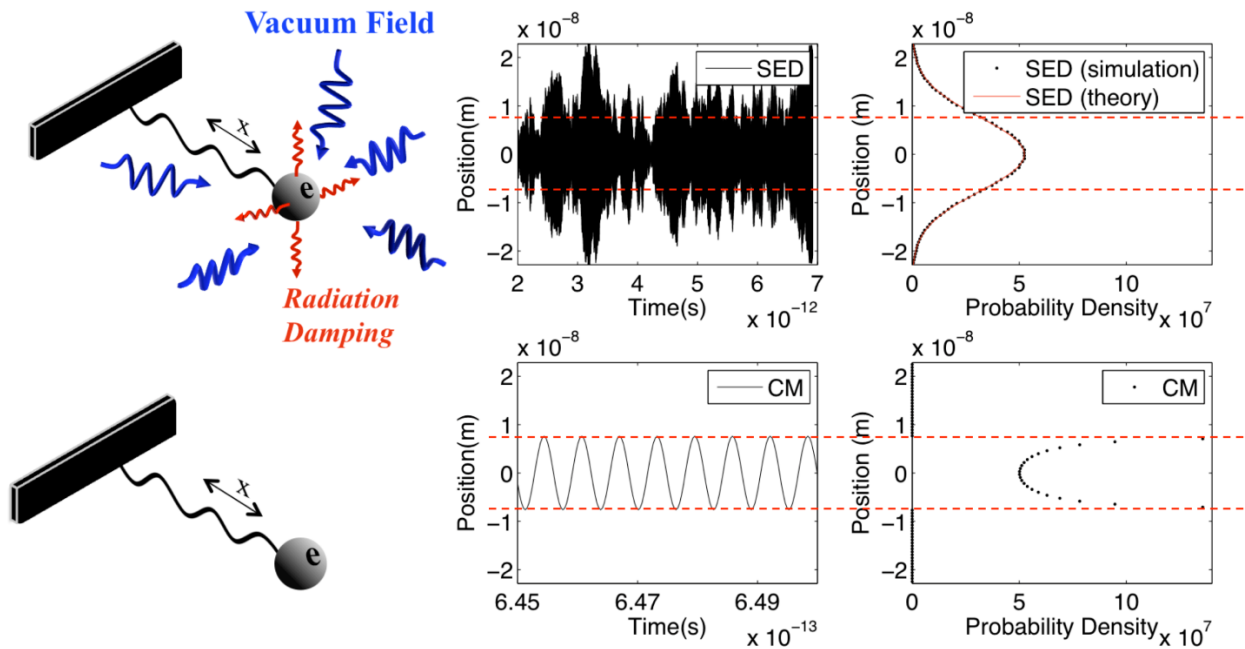

**Figure 9. Harmonic oscillators with and without the vacuum field. (Top) The SED harmonic oscillator undergoes an oscillatory motion with modulating oscillation amplitude. The oscillation amplitude modulates at the time scale of the coherence time and is responsible for the resulting Gaussian probability distribution. (Bottom) In the absence of the vacuum field or any external drive, a harmonic oscillator, that is initially displaced from equilibrium, performs simple harmonic oscillation with constant oscillation amplitude. The resulting probability distribution has peaks at the two turning points.**

The major challenge for simulating other physical systems, and in particular double slit diffraction, is to incorporate the appropriate SED vacuum field. A representative selection of vacuum field modes is thus the key to successful SED simulations. To push the limits of SED, and with the intention to exceed its validity regime we simulated a harmonic oscillator excited with a pulse of which the carrier frequency can be varied. The 1-D equation of motion in the x-direction used for the simulation is Newton equation for a charged particle damped by radiation and driven by the vacuum field and a pulsed field,

$$m\ddot{x} = -m\omega_0^2 x - m\Gamma \omega_0^2 \dot{x} + q\left[\left(E_p^{(x)} + E_{vac}^{(x)}\right) + \left(\vec{v} \times \left(\vec{B}_p + \vec{B}_{vac}\right)^{(x)}\right)\right]. \qquad 8$$

The field and parameters are described in ref [32]. The result of the simulation is shown in Fig. 10.

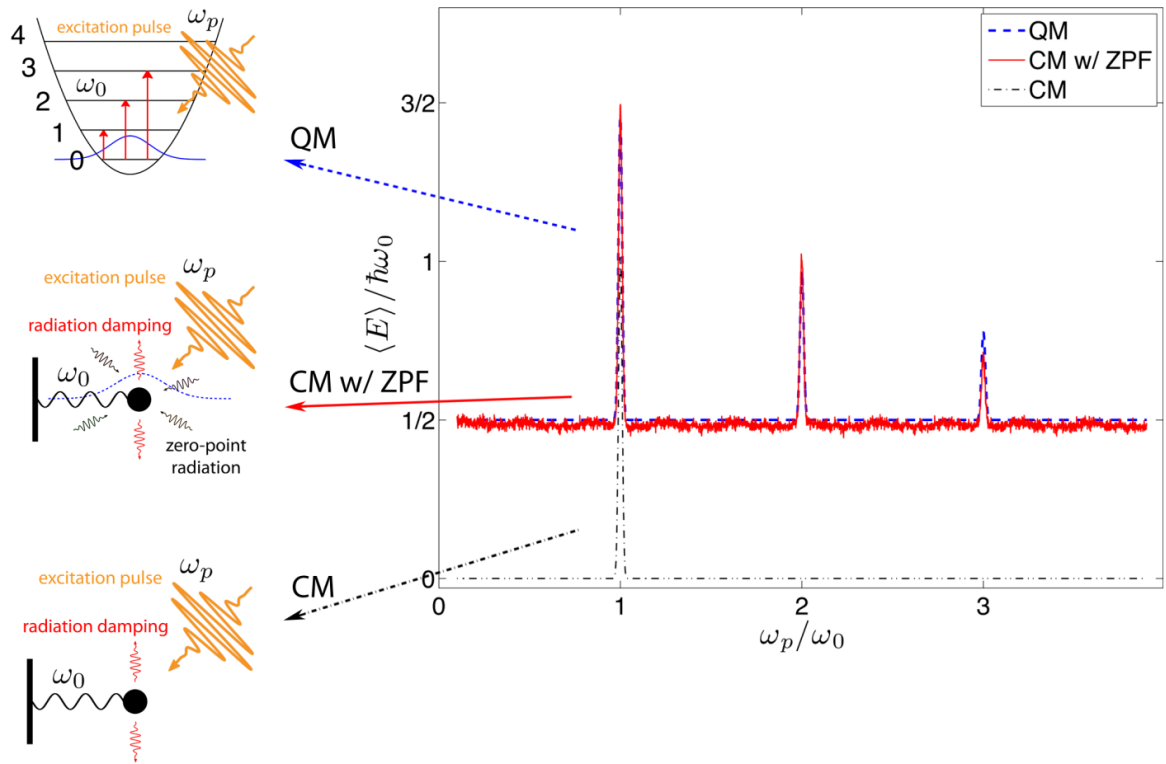

**Figure 10. Excitation spectra of harmonic oscillators for different theories. (Left) Schematics of harmonic oscillators are shown for quantum theory (left-top), for SED (left-middle), and for classical theory (left-bottom). The *red arrows* in the quantum system represents the one-step transition at different pulse frequencies $\omega_p$. (Right) The average value of energy $\langle E \rangle$ after excitation is plotted as a function of pulse frequency $\omega_p$. For the classical theory, the ensemble average is computed. For the quantum theory, the expectation value is computed. The classical oscillator in the vacuum field (red solid line) exhibits an excitation spectrum in agreement with the quantum result (blue dashed line). In the absence of the vacuum field, the classical oscillator has only one single resonance peak at the natural frequency $\omega_0$ (black dash-dot line). The excitation peak heights and the relative ratio are confirmed by classical and quantum perturbation theory** [32]**.**

The surprise was that the quantum mechanical result agrees well with the SED results, both for numerical simulation and perturbation theory. The $2\omega$ and $3\omega$ overtones, that are absent for a pure classical oscillator, can be explained by parametric excitation [32]. The more detailed agreement of the peak heights in the discrete excitation spectrum was not expected.

Returning to electron double slit diffraction, it may appear straightforward to run a simulation. However, a free electron can interact with all modes. Even ignoring infrared and ultraviolet divergences, an unbounded vacuum spectrum is hard to simulate. To sidestep this problem, Avendano is attempting to use a guiding equation based on a particle model to obtain electron trajectory in the vacuum [23]. The results are bundles of trajectories (Fig. 6) that may support diffraction peaks. At this point no probability distributions have been reported, and this work should also be considered unfinished.

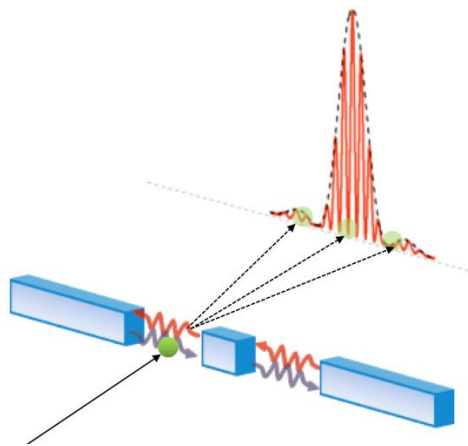

**Figure 11. Diffraction mechanism. (Left) Vacuum field modes are selected by the double-slit structure, which in turn guide the electron particle motion.**

It may appear that we can rule out SED as a viable theory as it may be real and local. Perhaps it cannot violate Bell's inequalities and should thus ultimately fail. This discussion, whether or not SED could be considered an emergent quantum theory, is interesting but not important for our present discussion. Even if SED is only an approximate theory with a limited validity range capable of reproducing only some quantum effects, the specific prediction that the vacuum field is responsible for diffraction is worth investigating in its own right. The alternative of using a full QED calculation for electron double slit diffraction, that includes the slits and the vacuum field, is technically more challenging and has never been done. The use of SED for this problem is then one of the options that is available, to address the main question raised in this paper: "By what physical process does the electron exchange momentum with the double slit?"

**Acknowledgements.** The authors acknowledge helpful discussions with Yves Couder, Ana Maria Cetto, Jaime Avendeno. This work is supported by the National Science Foundation under Grant No. 1306565.

**Bibliography.**
[1]   R. Bach, D. Pope, S.-H. Liou, and H. Batelaan, New J. Phys. **15**, 033018 (2013).
[2]   T. H. Boyer, Phys. Rev. D **11**, 790–808 (1975).
[3]   D. Hestenes, Found. Phys. **15**, 63–87 (1985).
[4]   A. F. Kracklauer, Found. Phys. Lett. **12**, 441–453 (2013).
[5]   G. Grössing, S. Fussy, J. Mesa Pascasio, and H. Schwabl, Ann. Phys. **327**, 421–437 (2012).
[6]   G. Cavalleri, F. Barbero, G. Bertazzi, E. Cesaroni, E. Tonni, L. Bosi, G. Spavieri, and G. T. Gillies, Front. Phys. China **5**, 107–122 (2009).
[7]   T. W. Marshall, Proc. R. Soc. Lond. Math. Phys. Eng. Sci. **276**, 475–491 (1963).
[8]   P. A. Schilpp, Ed., Albert Einstein, Philosopher-Scientist: The Library of Living Philosophers Volume VII, 3rd edition (Open Court, 1998).
[9]   J. A. Wheeler and W. H. Zurek, Quantum Theory and Measurement (Princeton University Press, 2014).
[10]  Feynman, Feynman Lectures on Physics, Volume III: Quantum Mechanics (CreateSpace Independent Publishing Platform, 2015).
[11]  W. K. Wootters and W. H. Zurek, Phys. Rev. D **19**, 473–484 (1979).
[12]  P. L. Kapitza and P. a. M. Dirac, Math. Proc. Camb. Philos. Soc. **29**, 297–300 (1933).
[13]  D. L. Freimund, K. Aflatooni, and H. Batelaan, Nature **413**, 142–143 (2001).
[14]  H. Batelaan, Rev. Mod. Phys. **79**, 929–941 (2007).
[15]  S. McGregor, W. C.-W. Huang, B. A. Shadwick, and H. Batelaan, Phys. Rev. A **92**, 023834 (2015).
[16]  C. J. Davisson and L. H. Germer, Proc. Natl. Acad. Sci. U. S. A. **14**, 317–322 (1928).
[17]  G. Gronniger, B. Barwick, H. Batelaan, T. Savas, D. Pritchard, and A. Cronin, Appl. Phys. Lett. **87**, 124104 (2005).
[18]  B. Barwick, G. Gronniger, L. Yuan, S.-H. Liou, and H. Batelaan, J. Appl. Phys. **100**, 074322 (2006).
[19]  L. Spruch, Phys. Today **39**, 37–45 (2008).
[20]  P. W. Milonni, Int. J. Theor. Phys. **22**, 323–328 (1983).
[21]  T. H. Boyer, Phys. Rev. A **7**, 1832–1840 (1973).
[22]  H. B. Wayne Cheng-Wei Huang, Found. Phys. **45** (2012).
[23]  L. de la Peña, A. M. Cetto, and A. Valdés Hernández, The Emerging Quantum (Springer International Publishing, Cham, 2015).
[24]  B. Hensen, H. Bernien, A. E. Dreau, A. Reiserer, N. Kalb, M. S. Blok, J. Ruitenberg, R. F. L. Vermeulen, R. N. Schouten, C. Abellan, W. Amaya, V. Pruneri, M. W. Mitchell, M. Markham, D. J. Twitchen, D. Elkouss, S. Wehner, T. H. Taminiau, and R. Hanson, Nature **526**, 682–686 (2015).
[25]  L. K. Shalm, E. Meyer-Scott, B. G. Christensen, P. Bierhorst, M. A. Wayne, M. J. Stevens, T. Gerrits, S. Glancy, D. R. Hamel, M. S. Allman, K. J. Coakley, S. D. Dyer, C. Hodge, A. E. Lita, V. B. Verma, C. Lambrocco, E. Tortorici, A. L. Migdall, Y. Zhang, et al., Phys. Rev. Lett. **115**, 250402 (2015).
[26]  M. Giustina, M. A. M. Versteegh, S. Wengerowsky, J. Handsteiner, A. Hochrainer, K. Phelan, F. Steinlechner, J. Kofler, J.-Å. Larsson, C. Abellán, W. Amaya, V. Pruneri, M. W. Mitchell, J. Beyer, T. Gerrits, A. E. Lita, L. K. Shalm, S. W. Nam, T. Scheidl, et al., Phys. Rev. Lett. **115**, 250401 (2015).
[27]  Y. Couder and E. Fort, Phys. Rev. Lett. **97**, 154101 (2006).
[28]  J. W. M. Bush, Phys. Today **68**, 47–53 (2015).
[29]  A. Andersen, J. Madsen, C. Reichelt, S. Rosenlund Ahl, B. Lautrup, C. Ellegaard, M. T. Levinsen, and T. Bohr, Phys. Rev. E **92**, 013006 (2015).
[30]  W. C.-W. Huang, H. Batelaan, W. C.-W. Huang, and H. Batelaan, J. Comput. Methods Phys. J. Comput. Methods Phys. **2013, 2013**, e308538 (2013).
[31]  T. H. Boyer, Sci. Am. **253**, 70–78 (1985).
[32]  W. C.-W. Huang and H. Batelaan, Found. Phys. **45**, 333–353 (2015).